\documentclass[submission, Proceedings]{SciPost}

\hypersetup{
    colorlinks,
    linkcolor={red!50!black},
    citecolor={blue!50!black},
    urlcolor={blue!80!black}
}

\usepackage[bitstream-charter]{mathdesign}
\DeclareSymbolFont{usualmathcal}{OMS}{cmsy}{m}{n}
\DeclareSymbolFontAlphabet{\mathcal}{usualmathcal}

\usepackage{graphics}
\usepackage{amsmath} 
\usepackage{amsfonts}
\usepackage{mathtools}
\usepackage{pstricks}
\usepackage[final]{pdfpages} 
\usepackage{ifpdf} 
\RequirePackage{ifpdf} 

\usepackage{color}
\definecolor{urlblue}{rgb}{0.2,0.4,0.7}
\definecolor{citegreen}{rgb}{0,0.6,0.2}
\definecolor{linkred}{rgb}{0.9,0.2,0.1}

\usepackage{caption} 
\usepackage{subcaption} 
\usepackage{autobreak}
\usepackage{marginnote}
\usepackage{multirow}

\usepackage{etoolbox} 
\usepackage{fixmath}
\usepackage{psfrag}
\usepackage{slashed}

\usepackage{notoccite} 

\newcommand{\NOdisplay}[1]{ }

\def\MSbar{\overline{\mathrm{MS}}}
\def\TR{{\displaystyle \mathrm{T}_{F}}}
\def\gJJ{\gamma_{{\scriptscriptstyle J}}}
\def\gFJ{\gamma_{{\scriptscriptstyle FJ}}}
\def\gFF{\gamma_{{\scriptscriptstyle F\tilde{F}}}}

\begin{document}

\begin{center}{\Large \textbf{
Renormalization of the flavor-singlet axial-vector current and its anomaly in QCD\\
}}\end{center}

\begin{center}
Long Chen\footnote{The speaker, together with T.~Ahmed and M.~Czakon, thank the organisers of RADCOR-LoopFest 2021.}
\end{center}

\begin{center}
Institut f\"ur Theoretische Teilchenphysik und Kosmologie, RWTH Aachen University,\\ Sommerfeldstr.~16, 52056 Aachen, Germany
\\
E-mail: longchen@physik.rwth-aachen.de
\end{center}



\definecolor{palegray}{gray}{0.95}
\begin{center}
\colorbox{palegray}{
  \begin{tabular}{rr}
  \begin{minipage}{0.1\textwidth}
    \includegraphics[width=35mm]{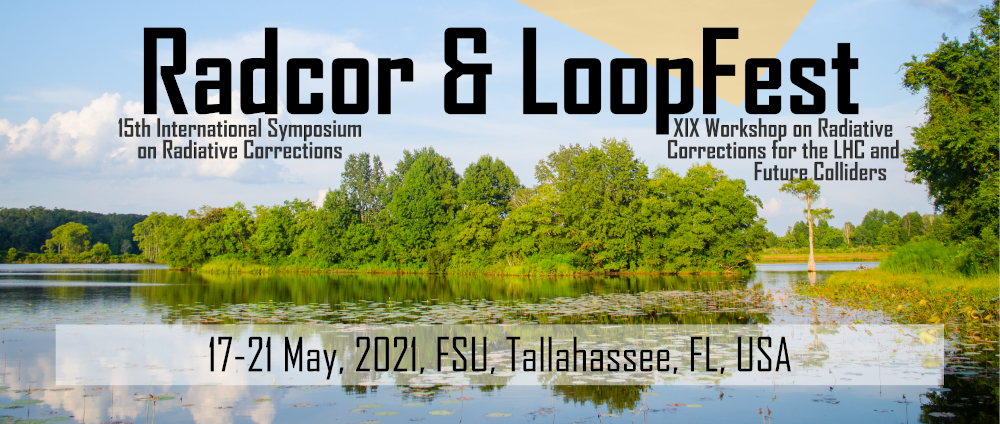}
  \end{minipage}
  &
  \begin{minipage}{0.85\textwidth}
    \begin{center}
    {\it 15th International Symposium on Radiative Corrections: \\Applications of Quantum Field Theory to Phenomenology,}\\
    {\it FSU, Tallahasse, FL, USA, 17-21 May 2021} \\
    \end{center}
  \end{minipage}
\end{tabular}
}
\end{center}

\section*{Abstract}
{\bf
The renormalization constant $Z_J$ of the flavor-singlet axial-vector current with a non-anticommuting $\gamma_5$ in dimensional regularization is determined to order $\alpha_s^3$ in QCD with massless quarks. 
In addition, the equality between the $\overline{\mathrm{MS}}$ renormalization constant $Z_{F\tilde{F}}$ associated with the operator $\big[F \tilde{F} \big]_{R}$ and that of $\alpha_s$ is verified explicitly to hold true at 4-loop order.
}

\vspace{10pt}
\noindent\rule{\textwidth}{1pt}
\tableofcontents\thispagestyle{fancy}
\noindent\rule{\textwidth}{1pt}
\vspace{10pt}

\section{Introduction}
\label{sec:intro}

Ever since the introduction of dimensional regularization (DR)~\cite{tHooft:1972tcz,Bollini:1972ui}, it has been recognized that special attention is required in the treatment of  $\gamma_5$, an intrinsically 4-dimensional object.
At the root of the issue is that a fully anticommuting $\gamma_5$ is algebraically incompatible with the Dirac algebra in D ($\neq 4$) dimensions, which on the other hand is essential for the concept of chirality of spinors in 4 dimensions and (non-anomalous) chiral symmetries in quantum field theory (QFT).
Furthermore, there is one well-known subtlety related to $\gamma_5$ and chiral symmetries in QFT, namely, that the flavor-singlet axial-vector current\footnote{Below we denote (flavor-singlet) axial-vector current simply by ``(singlet) axial current'' for brevity.}, defined with $\gamma_5$, exhibits a quantum anomaly, the axial or Adler-Bell-Jackiw (ABJ) anomaly~\cite{Adler:1969gk,Bell:1969ts}, in its divergence.
An anticommuting $\gamma_5$ together with the invariance of loop integrals under arbitrary loop-momentum shifts, however, leads to the absence of the axial anomaly, which can be easily checked at one-loop order which subsequently holds to all orders, due to the Adler-Bardeen theorem~\cite{Adler:1969er}. 

Despite the aforementioned difficulties, various $\gamma_5$ prescriptions in DR have been developed in the literature~\cite{tHooft:1972tcz,Akyeampong:1973xi,Breitenlohner:1977hr,Bardeen:1972vi,Chanowitz:1979zu,Gottlieb:1979ix,Ovrut:1981ne,Espriu:1982bw,Buras:1989xd,Kreimer:1989ke,Korner:1991sx,Larin:1991tj,Larin:1993tq,Jegerlehner:2000dz,Moch:2015usa,Zerf:2019ynn} over a span of nearly 50 years, albeit each with its own pros and cons.
In this talk, we discuss the calculation of the renormalization constant of the singlet axial current operator with a non-anticommuting $\gamma_5$~\cite{tHooft:1972tcz,Akyeampong:1973xi,Breitenlohner:1977hr} to the third  order in QCD, in the variant as prescribed in refs.~\cite{Larin:1991tj,Larin:1993tq}.~\footnote{Applications of a non-anticommuting $\gamma_5$ to the radiative corrections in electro-weak theory or even theories with beyond-Standard-Model interactions are more cumbersome, see e.g.~\cite{Barroso:1990ti,Jegerlehner:2000dz,BeluscaMaito:2020ala,BeluscaMaito:2021lnk}, and the result may depend on the details of the regularization prescription (see e.g.~\cite{Gnendiger:2017rfh}).}

\section{Renormalization of the singlet axial current}
\label{sec:aaop}

As summarized in refs.~\cite{Larin:1991tj,Larin:1993tq}, the properly renormalized singlet axial current in QCD with $n_f$ massless quarks with a non-anticommuting $\gamma_5$ can be written as 
\begin{eqnarray} 
\label{eq:J5uvr}
\left[ J^{\mu}_{5}\right]_{R} &=& \sum_{\psi_B} Z_J \, \bar{\psi}^{B}  \, \gamma^{\mu}\gamma_5 \, \psi^{B} \nonumber\\
&=& \sum_{\psi_B} Z^{f}_{5} \, Z^{ms}_{5} \,  \bar{\psi}^{B}  \, \frac{-i}{3!} \epsilon^{\mu\nu\rho\sigma} \gamma_{\nu} \gamma_{\rho} \gamma_{\sigma} \, \psi^{B} \,,
\end{eqnarray}
where $\psi^{B}$ denotes a bare quark field with mass dimension $(D-1)/2$ and the subscript $R$ at a square bracket denotes operator renormalization.~\footnote{In ref.~\cite{Ahmed:2021spj}, a factor $\mu^{4-D}$ in the mass scale $\mu$ of dimensional regularization was introduced in order for the mass dimension of the r.h.s.\ operator be equal to the canonical dimension of the l.h.s.\ in four dimensions.} 
The sum extends over all $n_f$ quark fields. 
Here and below $J^{\mu}_{5}$ denotes the bare flavor-singlet axial current.
It is known to renormalize multiplicatively~\cite{Adler:1969gk,Trueman:1979en}, as it is the only local composite current operator in the context of QCD that has the correct mass dimension and conserved quantum numbers (which are preserved under renormalization). 
The factor $Z_{J} \equiv Z^{f}_{5} \, Z^{ms}_{5}$ denotes the UV renormalization constant of the current,
conveniently parameterized as the product of a pure $\MSbar$-renormalization part $Z^{ms}_{5}$ and an additional finite renormalization  factor $Z^{f}_{5}$.
The latter is needed to restore the correct form of the axial Ward identity, namely, the all-order axial-anomaly equation~\cite{Adler:1969gk,Adler:1969er}, which reads in terms of renormalized local composite operators\footnote{When inserted into a Green's function, the time component of the derivative generates \textit{contact} terms in the respective Ward identity.}
\begin{eqnarray} 
\label{eq:ABJanomalyEQ}
\big[\partial_{\mu} J^{\mu}_{5} \big]_{R} = a_s\, n_f\, \TR \,  \big[F \tilde{F} \big]_{R}\,,
\end{eqnarray}
where $\TR=1/2$, $F \tilde{F} \equiv  - \epsilon^{\mu\nu\rho\sigma} F^a_{\mu\nu} F^a_{\rho\sigma} = \epsilon_{\mu\nu\rho\sigma} F^a_{\mu\nu} F^a_{\rho\sigma}$  denotes the contraction of the field strength tensor $F^a_{\mu\nu} = \partial_{\mu} A_{\nu}^{a} - \partial_{\nu} A_{\mu}^{a} + g_s \,  f^{abc} A_{\mu}^{b} A_{\nu}^{c}$ of the gluon field $A_\mu^a$ with its \textit{dual} form.
We use the shorthand notation $a_s \equiv \frac{\alpha_s}{4 \pi} = \frac{g_s^2}{16 \pi^2}$ for the QCD coupling, and $f^{abc}$ denotes the structure constants of the non-Abelian color group of QCD. 
In contrast to the l.h.s. of \eqref{eq:ABJanomalyEQ}, the renormalization of the axial-anomaly operator $F \tilde{F}$ is not strictly multiplicative (as known from ref.~\cite{Adler:1969gk}), but involves mixing with the divergence of the axial current operator~\cite{Espriu:1982bw,Breitenlohner:1983pi}, 
\begin{eqnarray} 
\label{eq:FFuvr}
\big[F \tilde{F} \big]_{R} = Z_{F\tilde{F}} \, \big[F \tilde{F} \big]_{B} \,+\, 
 Z_{FJ} \, \big[\partial_{\mu} J^{\mu}_{5} \big]_{B} \,,
\end{eqnarray}
where the subscript $B$ implies that the fields in the local composite operators are bare.
In the computation of the matrix elements of the axial-anomaly operator $F \tilde{F}$, we employ its equivalent form in terms of the divergence of the Chern-Simons current $K^{\mu}$, namely 
\begin{eqnarray} 
\label{eq:Kcurrent}
F \tilde{F}  &=& \partial_{\mu} K^{\mu} \nonumber\\
 &=& \partial_{\mu} \left(-4 \,\epsilon^{\mu\nu\rho\sigma} \,\left(A_{\nu}^{a} \partial_{\rho} A_{\sigma}^{a} \,+\, g_s\,\frac{1}{3} f^{abc} A_{\nu}^{a} A_{\rho}^{b} A_{\sigma}^{c} \right) \right)\,,
\end{eqnarray}
where, unlike $J^{\mu}_{5}$, the current $K^{\mu} \equiv -4 \, \epsilon^{\mu\nu\rho\sigma} \,\left(A_{\nu}^{a} \partial_{\rho} A_{\sigma}^{a} \,+\, g_s\,\frac{1}{3} f^{abc} A_{\nu}^{a} A_{\rho}^{b} A_{\sigma}^{c} \right)$ is not a gauge-invariant object~\cite{Espriu:1982bw,Breitenlohner:1983pi}.
The Feynman rules for the r.h.s. of eq.~\eqref{eq:ABJanomalyEQ} used in our calculation are directly based on eq.~\eqref{eq:Kcurrent}.

The renormalization of the operators $\partial_{\mu} J^{\mu}_{5}$ and $F \tilde{F}$ specified, respectively, in eq.~\eqref{eq:J5uvr} and eq.~\eqref{eq:FFuvr} can be arranged into the following matrix form 
\begin{eqnarray}
\label{eq:Zsmatrix}
\begin{pmatrix}
\big[\partial_{\mu} J^{\mu}_{5} \big]_{R}\\
\big[F \tilde{F}\big]_{R}
\end{pmatrix}
= 
\begin{pmatrix}
Z_{J} &  0 \\
Z_{FJ} &  Z_{F\tilde{F}}
\end{pmatrix}
\cdot 
\begin{pmatrix}
\big[\partial_{\mu} J^{\mu}_{5} \big]_{B}\\
\big[F \tilde{F}\big]_{B}
\end{pmatrix}\,.
\end{eqnarray}
The matrix of anomalous dimensions of these two renormalized operators is defined by 
\begin{eqnarray}
\label{eq:AMDmatrix}
\frac{\mathrm{d}}{\mathrm{d}\, \ln \mu^2}\,
\begin{pmatrix}
\big[\partial_{\mu} J^{\mu}_{5} \big]_{R}\\
\big[F \tilde{F}\big]_{R}
\end{pmatrix}
= 
\begin{pmatrix}
\gJJ &  0 \\
\gFJ &  \gFF
\end{pmatrix}
\cdot 
\begin{pmatrix}
\big[\partial_{\mu} J^{\mu}_{5} \big]_{R}\\
\big[F \tilde{F}\big]_{R}\,.
\end{pmatrix}
\end{eqnarray}

\section{The axial-anomaly projector}
\label{sec:proj}

\begin{figure}[htbp]
\begin{center}
\includegraphics[scale=0.8]{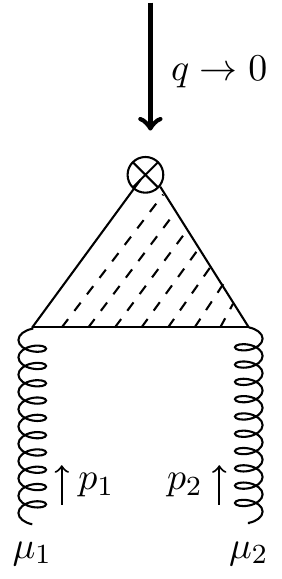}
\caption{The kinematic configuration selected for the computations of matrix elements of the operators appearing in the axial-anomaly equation.}
\label{fig:kinematics}
\end{center}
\end{figure}

We choose to determine the renormalization constant $Z_{J} \equiv Z^{f}_{5} \, Z^{ms}_{5}$ of the singlet axial current to $\mathcal{O}(a^3_s)$, and $Z_{F\tilde{F}}$ to $\mathcal{O}(a^4_s)$, in DR by computing matrix elements of operators appearing in the axial-anomaly equation between the vacuum and a pair of off-shell gluons evaluated at a specifically chosen single-scale kinematic configuration~\cite{Bos:1992nd,Larin:1993tq}, as illustrated in figure~\ref{fig:kinematics}.
Let us denote by $\langle 0| \hat{\mathrm{T}}\left[ J_{5}^{\mu}(y) \, A_a^{\mu_1}(x_1) \, A_a^{\mu_2}(x_2) \right] |0 \rangle |_{\mathrm{amp}}$ the amputated one-particle irreducible (1PI) vacuum expectation value of the time-ordered covariant product of the (singlet) axial current and two gluon fields in coordinate space with open Lorentz indices. 
Subsequently, we introduce the following rank-3 matrix element $\Gamma^{\mu \mu_1 \mu_2}_{lhs}(p_1, p_2)$ in momentum space, defined by
\begin{eqnarray}
\label{eq:Glhs1PI}
\Gamma^{\mu \mu_1 \mu_2}_{lhs}(p_1, p_2) \equiv 
\int d^4 x  d^4 y \, e^{-i p_1 \cdot x - i q \cdot y }\,  \langle 0| \hat{\mathrm{T}}\left[ J_{5}^{\mu}(y) \, A_a^{\mu_1}(x) \, A_a^{\mu_2}(0) \right] |0 \rangle |_{\mathrm{amp}} 
\end{eqnarray}
where translation invariance 
has been used to shift the coordinate of one gluon field to the origin, and the resulting total momentum conservation factor, ensuring $p_2 = -q - p_1$, is implicit. 

Rather than performing the projection for the axial anomaly literally as devised in eq.~(19) of ref.~\cite{Larin:1993tq} where a derivative w.r.t the momentum $q$ is taken before going to the limit $q_\mu \rightarrow 0$~(see also ref.~\cite{Bos:1992nd}), we use instead the following projector 
\begin{eqnarray}
\label{eq:anomalyprojector}
\mathcal{P}_{\mu \mu_1 \mu_2} = -\frac{1}{6 \, p_1 \cdot p_1} \, \epsilon_{\mu\mu_1\mu_2\nu}\, p_1^{\nu} \,,
\end{eqnarray}
and directly project $\Gamma^{\mu \mu_1 \mu_2}(p_1, -p_1)$ onto this structure right at $q = 0$ (i.e.~$p_2=-p_1$). 
This leads to the scalar mass-dimensionless matrix element
\begin{eqnarray}
\label{eq:smeL}
\mathcal{M}_{lhs} =  \mathcal{P}_{\mu \mu_1 \mu_2} \,  \Gamma^{\mu \mu_1 \mu_2}_{lhs}(p_1, -p_1) \,.
\end{eqnarray}

Although, at first sight, the projector $\mathcal{P}_{\mu \mu_1 \mu_2}$ in eq.(\ref{eq:anomalyprojector}) does not seem to have anything to do with the divergence or anomaly of the axial current, it can be shown that with the appropriate regularity condition it is indeed equivalent to the operation devised for projecting out the anomaly in eq.(19) of ref.~\cite{Larin:1993tq}; see the Appendix of ref.~\cite{Ahmed:2021spj}.
Another useful way to appreciate the connection between this projector and the anomaly of the axial current is by examining the form factor decomposition of the matrix element $\Gamma^{\mu \mu_1 \mu_2}_{lhs}(p_1, p_2)$ in eq.(\ref{eq:Glhs1PI}).
It involves the following 3 tensor structures~\cite{Rosenberg:1962pp}, 
\begin{eqnarray}
\label{eq:FFDec}
\Gamma^{\mu \mu_1 \mu_2}_{lhs}(p_1, p_2) &=&
F_1 \, \epsilon^{\mu\,\mu_1\,\mu_2\, (p_2-p_1)} \nonumber\\
&+& F_2 \, \big(p_1^{\mu_1} \epsilon^{\mu\,\mu_2\,p_1\, p_2} - p_2^{\mu_2} \epsilon^{\mu\,\mu_1\,p_1\, p_2} \big)  \nonumber\\
&+& F_3 \, \big(p_1^{\mu_2} \epsilon^{\mu\,\mu_1\,p_1\, p_2} - p_2^{\mu_1} \epsilon^{\mu\,\mu_2\,p_1\, p_2} \big)\,
\end{eqnarray}
each of which is respectively parity-odd and Bose-symmetric w.r.t the two external gluons.~\footnote{There are 6 linearly independent rank-3 Lorentz tensor structures that one can compose in 4 dimensions out of the two momenta $p_1, p_2$ and one Levi-Civita tensor. They can be reduced to the three included in eq.(\ref{eq:FFDec}) upon imposing Bose symmetry and parity odd conditions. We note further that the external gauge bosons are not required to be transversal in this decomposition.} 
The form factors $F_{1,2,3}$ are functions of the Lorentz invariants made out of $p_1, p_2$.
With this parameterization, the contraction of this tensor with the sum of $p_1$ and $p_2$ receives contribution from just $F_1$:
\begin{eqnarray*}
-q_{\mu}\Gamma^{\mu \mu_1 \mu_2}_{lhs}(p_1, p_2) &=& 2 F_1 \, \epsilon^{\mu_1\,\mu_2\, p_1\,p_2}\,, 
\end{eqnarray*} 
which corresponds to the divergence of the axial current.
With on-shell $p_1$ and $p_2$, the $F_1$ is known to vanish at $q^2=0$ in an abelian theory with all fermions massive,  due to a low-energy theorem ~\cite{Sutherland:1967vf}.
It is also clear from eq.(\ref{eq:FFDec}) that there is only $\epsilon^{\mu\,\mu_1\,\mu_2\, (p_2-p_1)}$ surviving at the chosen kinematics $q=0, p_2 = - p_1$.
Consequently, the projector $\mathcal{P}_{\mu \mu_1 \mu_2}$ projects out the form factor in front of this unique structure, which is not vanishing in the limit $q=0$ with off-shell $p_1$ (and $p_2$).

Strictly speaking, the squared norm of the Lorentz structure $\epsilon_{\mu\mu_1\mu_2\nu} p_1^{\nu}$ is equal to (6-11$D$+6$D^2$-$D^3\,$)$p_1\cdot p_1$ with all Lorentz indices of the space-time metric tensors resulting from contracting a pair of Levi-Civita tensors
taken to be $D$-dimensional, but it is known~\cite{Chen:2019wyb,Ahmed:2019udm} that the parameter $D$ here can be safely set to 4 consistently throughout the computation in DR without problem~\footnote{This has already been employed extensively in ref.~\cite{Chen:2019wyb} for devising projectors to project out amplitudes directly both in the linear polarization and helicity basis, and along a similar line of thinking it was used for conventional form factor projectors in ref.~\cite{Ahmed:2019udm}. In physical applications the number of these projections needed are equal to the number of independent helicity amplitudes in 4 dimensions.}.
In fact, it is also convenient to do so in our computational set-up as the projection of $\mathcal{M}_{lhs}$ is separated from the evaluation and substitution of each individual loop integral therein.

The scalar quantity $\mathcal{M}_{lhs}$ is projected out and evaluated right at the limit of zero momentum $q = -(p_1 + p_2)$ flowing through the inserted operator $J_{5}^{\mu}(y)$, but with off-shell gluon momentum $p_1^2 \neq 0$, for the sake of the following:  
\begin{itemize}
\item 
the possible IR divergences in $\Gamma^{\mu \mu_1 \mu_2}_{lhs}(p_1, p_2)$ are nullified and all the poles in $\epsilon$ evaluated with off-shell momentum $p_1$ are of UV origin owing to the IR-rearrangement~\cite{Vladimirov:1979zm};
\item 
all loop integrals involved are reduced to massless propagator-type integrals which are well studied and known numerically~\cite{Smirnov:2010hd} as well as analytically~\cite{Baikov:2010hf,Lee:2011jt} to 4-loop order.
\end{itemize}
Because $p_1^2\neq 0$ is the only physical scale involved in the matrix elements under consideration, we set $p_1^2 = 1$ from the outset in our computations without loss of generality.

The very same projector $\mathcal{P}_{\mu \mu_1 \mu_2}$ is also used in extracting a scalar quantity from the matrix element of $\big[F \tilde{F} \big]_{R}$ between the vacuum and the same external (off-shell) gluon state.
In order to be able to apply eq.~(\ref{eq:anomalyprojector}), it is crucial to use the equivalent form of the axial-anomaly operator $F \tilde{F}  = \partial_{\mu} K^{\mu}$ in terms of the divergence of the current $K^{\mu}$.
To be specific, we define in analogy to eq.~\eqref{eq:Glhs1PI}:
\begin{eqnarray}
\label{eq:Grhs1PI}
\Gamma^{\mu \mu_1 \mu_2}_{rhs}(p_1, p_2) \equiv 
\int d^4 x  d^4 y \, e^{-i p_1 \cdot x - i q \cdot y }\,  \langle 0| \hat{\mathrm{T}}\left[ K^{\mu}(y) \, A_a^{\mu_1}(x) \, A_a^{\mu_2}(0) \right] |0 \rangle |_{\mathrm{amp}} \,.
\end{eqnarray}
Subsequently, one contracts $\Gamma^{\mu \mu_1 \mu_2}_{rhs}(p_1, -p_1)$ with $\mathcal{P}_{\mu \mu_1 \mu_2}$, yielding a r.h.s. counterpart to eq.~\eqref{eq:smeL}:  
\begin{eqnarray}
\label{eq:smeR}
\mathcal{M}_{rhs} =  \mathcal{P}_{\mu \mu_1 \mu_2} \,  \Gamma^{\mu \mu_1 \mu_2}_{rhs}(p_1, -p_1) \,.
\end{eqnarray}
The perturbative corrections to $\mathcal{M}_{rhs}$ in terms of Feynman diagrams should be computed using the Feynman rules derived from eq.~\eqref{eq:Kcurrent}.

\section{Results and discussions}

We compute the pertubative QCD corrections to $\mathcal{M}_{lhs}$ and $\mathcal{M}_{rhs}$ in terms of Feynman diagrams, which are manipulated in the usual way.
We refer to ref.~\cite{Ahmed:2021spj} for details on the technical aspects of the computation.
Whereas the off-shell matrix elements $\mathcal{M}_{lhs}$ and $\mathcal{M}_{rhs}$ depend on the gluon-field gauge-fixing parameter $\xi$, the renormalization constant $Z_J \equiv Z^{ms}_{5} \, Z^{f}_{5}$ of the gauge-invariant axial-current operator $J_5^{\mu}$ is independent of $\xi$.
The $\MSbar$ part of the renormalization constant $Z^{ms}_{5}$ can be extracted based on the UV divergences remaining in the $\mathcal{M}_{lhs}$ after performing the renormalization of the external (off-shell) gluon fields and the QCD coupling $a_s$, as well as the renormalization of the covariant-gauge fixing parameter $\xi$.
In the similar way, the $\MSbar$ renormalization constant $Z_{F\tilde{F}}$ can also be determined from the result for $\mathcal{M}_{rhs}$, given the knowledge of all other renormalization constants to the respective perturbative orders in need.
The finite renormalization constant $Z^{f}_{5}$ is determined by demanding the equality between the fully renormalized $\mathcal{M}_{lhs}$ and $\mathcal{M}_{rhs}$, originating from eq.~\eqref{eq:ABJanomalyEQ}.
It reads~\cite{Ahmed:2021spj}: 
\begin{align}
\label{eq:zf5}
\begin{autobreak}
Z^{f}_{5} =
    1    
    +a_s \Big\{ -4 {C_F} \Big\}   
    + a_s^2 \Big\{ {C_A} {C_F}\Big( -\frac{107 }{9}\Big)+ C_F^2 \Big( 22 \Big) + {C_F} {n_f} \Big( \frac{31 }{18} \Big)  \Big\}       
    + a_s^3 \Big\{ C_A^2 {C_F} \Big(56 \zeta_3-\frac{2147}{27}\Big)    
    + {C_A} C_F^2 \Big(\frac{5834}{27}-160 \zeta_3\Big)    
    + {C_A} {C_F} {n_f} \Big(\frac{110 }{3}\zeta_3-\frac{133}{81}\Big)   
   +C_F^3 \Big(96 \zeta_3-\frac{370}{3}\Big)   
   +C_F^2 {n_f} \Big(\frac{497}{54}-\frac{104 }{3}\zeta_3\Big)   
   + {C_F} n_f^2 \Big( \frac{316 }{81}  \Big) \Big\}\,.
\end{autobreak}
\end{align}
The definition of the quadratic Casimir color constants is as usual: $C_A = N_c \,, \, C_F = (N_c^2 - 1)/(2 N_c) \,$ along with the color-trace normalization factor $\TR = {1}/{2}$.
We note that the explicit perturbative result in eq.(\ref{eq:zf5}) is given in terms of the usual $\MSbar$-renormalized QCD coupling $a_s$ with $n_f$ quark flavors.
The first two orders of eq.~\eqref{eq:zf5} agree with ref.~\cite{Larin:1993tq}, while the third order terms are our new result.
~\\

We note that the difference between $Z^{f}_{5}$ in eq.~\eqref{eq:zf5} and the additional finite renormalization constant for the non-singlet axial current computed to $\mathcal{O}(a_s^3)$ in ref.~\cite{Larin:1991tj} starts at $\mathcal{O}(a_s^2)$, and is proportional to $n_f\, C_F$ just like their $\MSbar$ counterparts. 
There is actually a quite interesting point related to this, which we elaborate below, given the common practice of splitting the QCD corrections to the axial part of the quark form factors into the so-called non-singlet and singlet contribution (see, e.g.,~\cite{Bernreuther:2005rw,Gehrmann:2021ahy,Chen:2021rft}.)
It starts with the following question: if there are $n_f$ flavors of quarks active in the QCD Lagrangian, what should be the renormalized form of a manually separated axial current component with one particular quark flavor.
Namely, instead of the complete singlet axial current in eq.(\ref{eq:J5uvr}), we consider the renormalization of an individual component $\bar{\psi}_{q} \, \gamma^{\mu}\gamma_5 \, \psi_{q}$ with the subscript $q$ denoting the quark flavor in QCD with $n_f$ quarks active.
Apparently, by definition, one should have 
\begin{eqnarray}
\label{eq:J5uvr_split}
\sum_{q=1}^{n_f} \, \big[J^{\mu}_{5,q}\big]_R = \big[J^{\mu}_{5}\big]_R
\end{eqnarray}
with $\big[J^{\mu}_{5}\big]_R$ the renormalized complete singlet axial current in eq.(\ref{eq:J5uvr}) and $\big[J^{\mu}_{5,q}\big]_R$ denoting an individual component of a given quark flavor $q$.
The point is that it would be incorrect to make the naive identification of  $\big[J^{\mu}_{5,q}\big]_R$ as $Z_J \, \bar{\psi}^{B}_q  \, \gamma^{\mu}\gamma_5 \, \psi^{B}_q$, the latter of which is actually not UV finite, although the condition~(\ref{eq:J5uvr_split}) would be trivially fulfilled.
Instead, the correct form should read 
\begin{eqnarray}
\label{eq:J5quvr}
\big[J^{\mu}_{5,q}\big]_R &=& 
Z_{ns}\, \bar{\psi}^{B}_{q}  \, \gamma^{\mu}\gamma_5 \, \psi^{B}_q \,+\, Z_s\,\sum_{i=1}^{n_f} \, \bar{\psi}^{B}_{i}  \, \gamma^{\mu}\gamma_5 \, \psi^{B}_i \nonumber\\
&=& \big( Z_{ns}\, + \, Z_s\, \big)\, \bar{\psi}^{B}_{q}  \, \gamma^{\mu}\gamma_5 \, \psi^{B}_q \,+\, 
Z_s\, \sum_{i=1, i \neq q}^{n_f} \, \bar{\psi}^{B}_{i}  \, \gamma^{\mu}\gamma_5 \, \psi^{B}_i \,,
\end{eqnarray}
with $Z_s \equiv \frac{1}{n_f} \big( Z_J - Z_{ns} \big)$ and $Z_{ns}$ denoting the full renormalization constant for the non-singlet axial current operator (i.e.,~including the non-$\MSbar$ finite piece). 
Owing to the aforementioned feature regarding the difference between the non-singlet and singlet axial-current renormalization constants, the so-defined $Z_s$ starts from $\mathcal{O}(a_s^2)$ and is free of $n_f$ in the denominator. 
It is interesting to observe that in the explicit renormalized form $\big[J^{\mu}_{5,q}\big]_R$ of an axial-current component of a given quark flavor $q$, there appears some mixing terms made out of the fields of the remaining quark flavors.
It is straightforward to see that the condition~(\ref{eq:J5uvr_split}) is satisfied by the formula~(\ref{eq:J5quvr}). 
Consequently, one can also derive the RG equation for $\big[J^{\mu}_{5,q}\big]_R$ based on eq.(\ref{eq:J5quvr}), which reads
\begin{eqnarray}
\label{eq:RGEssac}
\mu^2\frac{\mathrm{d} }{\mathrm{d} \mu^2} \big[J^{\mu}_{5,q}\big]_R 
&=&  \gamma_s \, \big[J^{\mu}_{5}\big]_R\,,
\end{eqnarray}
where $\gamma_s$ is defined by $\mu^2\frac{\mathrm{d} Z_s}{\mathrm{d} \mu^2}  = \gamma_s \big(Z_{ns} + n_f \, Z_{s}\big)$.
We note that it is the renormalized complete axial current $\big[J^{\mu}_{5}\big]_R$ that appears in the r.h.s. of eq.(\ref{eq:RGEssac}).
~\\

The matrix element $\mathcal{M}_{rhs}$ is only needed to 3-loop order in extracting the result for $Z^{f}_{5}$ in eq.~\eqref{eq:zf5}.
With its 4-loop expression at hand, we have determined the $Z_{F\tilde{F}}$ constant to $\mathcal{O}(a^4_s)$, where this renormalization constant is the only one involved to $\mathcal{O}(a^4_s)$ (apart from the known QCD field renormalization constant $Z_3$).
Subsequently, we have verified explicitly that 
\begin{eqnarray}
\label{eq:ZFFidentity}
Z_{F\tilde{F}} = Z_{a_s}  
\end{eqnarray}
holds true at $\mathcal{O}(a^4_s)$.
The Abelian version of eq.~\eqref{eq:ZFFidentity} was known already since the early reference~\cite{Adler:1969gk}.
This relation was first verified explicitly in QCD to $\mathcal{O}(a_s)$ in ref.~\cite{Espriu:1982bw}, to $\mathcal{O}(a^2_s)$ in ref.~\cite{Bos:1992nd,Larin:1993tq} and subsequently to $\mathcal{O}(a^3_s)$ in refs.~\cite{Zoller:2013ixa,Ahmed:2015qpa}, and now finally to $\mathcal{O}(a^4_s)$ where non-quadratic color Casimirs start to appear. 
This equality was shown in ref.~\cite{Breitenlohner:1983pi} to be needed to ensure the extension of the Adler-Bardeen theorem to a non-abelian gauge theory. 
It was further claimed to hold true in the same reference by showing that, apart from the mixing with the divergence of the axial-current operator, the operator $\hat{a}_s \, \big[F\tilde{F}\big]$ does not need any multiplicative renormalization in a gauge-invariant regularization scheme using the background field method~\cite{Abbott:1980hw}.
Very recently a proof of the absence of a (divergent) multiplicative renormalization of the axial-anomaly operator in dimensionally regularized QCD is completed in ref.~\cite{Luscher:2021bog}.
~\\

Subjecting both sides of the axial-anomaly equation~\eqref{eq:ABJanomalyEQ} to the logarithmic derivative $\mu^2 \frac{\mathrm{d}\, }{\mathrm{d}\, \mu^2}$, one obtains~\cite{Larin:1993tq}  
\begin{eqnarray}
\label{eq:ABJanomalyEQsAMD}
\left(\gJJ - n_f \, \TR \, a_s\, \gFJ - \gFF - \beta \right) \big[\partial_{\mu} J^{\mu}_{5}\big]_{R} = 0\,,
\end{eqnarray}
which actually holds only in the 4-dimensional limit $\epsilon = 0$.
The equality between the $\MSbar$ renormalization constants, $Z_{F\tilde{F}}$ and $Z_{a_s}$, explicitly verified to 4-loop order, implies the equality $\gFF = -\beta$ with $\beta \equiv - \mu^2\frac{\mathrm{d} \ln Z_{a_s}}{\mathrm{d} \mu^2}$.
Clearly, eq.~\eqref{eq:ABJanomalyEQsAMD}, which is a natural consequence of the all-order axial-anomaly equation~\eqref{eq:ABJanomalyEQ}, combined with $\gFF = -\beta$ necessarily implies that 
\begin{eqnarray}
\label{eq:AMDsEQ}
\gJJ = n_f \, \TR \, a_s\, \gFJ 
\end{eqnarray}
must hold true, albeit only in the limit $\epsilon = 0$.
Reformulating the axial-anomaly equation~\eqref{eq:ABJanomalyEQ} in terms of the bare fields, one arrives at 
\begin{eqnarray}
\label{eq:ABJanomalyEQbare}
\big(\, Z_{J} - n_f \, \TR \, a_s\, Z_{FJ} \big) \big[\partial_{\mu} J^{\mu}_{5}\big]_{B} = \hat{a}_s \, n_f \, \TR\, \big[F \tilde{F}\big]_{B}\,.
\end{eqnarray}
The RG-invariance of the operator on the r.h.s. of eq.~(\ref{eq:ABJanomalyEQbare}) is exact in $D$ dimensions as long as the Levi-Civita tensor in the definition of $F\tilde{F}$ is algebraically consistently defined. 
The RG-invariance of the l.h.s.~operator, on the other hand, is only conditionally true in the 4-dimensional limit. 
This reflects the fact that eq.~\eqref{eq:ABJanomalyEQ}, as well as eq.~\eqref{eq:ABJanomalyEQbare}, should be regarded as a relation valid in the 4-dimensional limit, rather than a $D$-dimensional identity.
It was shown by explicit calculations in ref.~\cite{Espriu:1982bw} that the coefficient $\big(\, Z_{J} - n_f \, \TR \, a_s\, Z_{FJ} \big)$ in front of the bare axial current $\big[J^{\mu}_{5}\big]_{B}$ on the l.h.s of eq.~\eqref{eq:ABJanomalyEQbare} reduces to unity up to $\mathcal{O}(a_s^2)$ in QCD with the $\gamma_5$ prescription of ref~\cite{Chanowitz:1979zu}.
Apparently, with a non-anticommuting $\gamma_5$ prescription, this coefficient is no longer unity, but the axial current in the l.h.s. of eq.~(\ref{eq:ABJanomalyEQbare}) remains RG-invariant, albeit only in the 4-dimensional limit.

\section{Conclusion}
\label{sec:conc}

We have extended the knowledge of the renormalization constant $Z_J$, and in particular, of the finite non-$\MSbar$ factor $Z_5^f$, of the flavor-singlet axial-vector current regularized with a non-anticommuting $\gamma_5$ to $\mathcal{O}(\alpha_s^3)$ in QCD through computations of matrix elements of operators appearing in the axial-anomaly equation $\big[\partial_{\mu} J^{\mu}_{5} \big]_{R} = a_s\, n_f\, \TR \,  \big[F \tilde{F} \big]_{R}$ between the vacuum and a pair of (off-shell) gluons to 4-loop order.  
We have verified explicitly up to 4-loop order the equality between the $\MSbar$ renormalization constant $Z_{F\tilde{F}}$ associated with the operator $\big[F \tilde{F} \big]_{R}$ and that of $\alpha_s$, which is recently proved to all orders in dimensionally regularized QCD.

\bibliography{Z5singlet} 

\nolinenumbers

\end{document}